\documentclass[twocolumn,english,prb,epsfig,rotate,showpacs,aps]{revtex4-2}
\usepackage[T1]{fontenc}
\usepackage[latin9]{inputenc}
\usepackage{color}
\usepackage{epsfig}
\usepackage{verbatim}
\usepackage{amsmath}
\usepackage{amsthm}
\usepackage{amssymb}
\usepackage{graphicx}
\usepackage{esint}
\usepackage{microtype}
\usepackage{dcolumn}
\usepackage{bm}
\usepackage{amsfonts}
\usepackage[table]{xcolor} %
\usepackage{subfigure}

\newcommand{\blue}{\color{black}}
\newcommand{\green}{\color{black}}
\newcommand{\bk}{\mathbf{k}}
\usepackage{titlesec}
\titlespacing*{\section}{0pt}{2ex plus 1ex minus .2ex}{2ex plus .2ex}
\makeatletter
\usepackage{babel}
\usepackage{braket}

\makeatother

\usepackage{babel}
\begin{document}
\renewcommand{\figurename}{Fig.}
\title{Robust boundary Luttinger surfaces in topological band structures}
\author{Kai Chen }
\affiliation{Department of Physics and Texas Center for Superconductivity, University
of Houston, Houston, TX 77204}
\author{Pavan Hosur}
\affiliation{Department of Physics and Texas Center for Superconductivity, University
of Houston, Houston, TX 77204}
\date{\today}
\begin{abstract}

The standard paradigm of topological phases posits that two phases with identical symmetries are separated by a bulk phase transition, while symmetry breaking provides a path in parameter space that allows adiabatic connection between the phases. Typically, if symmetry is broken only at the boundary, topological surface states become gapped, and single-particle surface properties no longer distinguish between the two phases. In this work, we challenge this expectation. We demonstrate that the single-particle surface Green's function contains zeros, or "Luttinger surfaces," which maintain the same bulk-boundary correspondence as topological surface states. Remarkably, these Luttinger surfaces persist under symmetry-breaking perturbations that destroy the surface states. Moreover, we point out that low-energy and surface theories, often used synonymously in discussions of (gapped) topological matter, are actually different, with the difference captured by the Luttinger surfaces.
\end{abstract}
\maketitle
\section{Introduction}
Bulk-boundary correspondence is
a fundamental notion that underpins the standard paradigm of topological
band structures. According to this principle, topological materials
host robust, gapless surface states that uniquely identify the bulk
band topology \citep{hasan2010colloquium,qi2011topological,volovik2003universe,armitage2018weyl,graf2013bulk,rhim2018unified}.
In fact, in practice, the surface states invariably serve as the most direct experimental smoking guns of the topological phase. When the bulk undergoes a topological phase transition, the surface also experiences reconstruction, appearance
or disappearance of states -- in short, a Lifshitz transition \cite{volovik2017topological} --
in accordance with the bulk-boundary correspondence. If the topological
phase is protected by a symmetry, breaking the symmetry generically
destroys the surface states and unlocks a path in parameter space
that allows one to adiabatically connect topologically distinct phases.
In particular, if the symmetry is broken only on the surface, the
bulk topology is well-defined but the surface generically remains
gapped along this path. As a result, single-particle physics of the
surface is expected to vary smoothly across the bulk phase transition
and be qualitatively blind to the bulk topology.

The single-particle Green's function \textbf{$G(\omega,\mathbf{k})$
}is a fundamental tool in physics, crucial for characterizing the
spectral properties of single-particle excitations as well as describing
their response to external perturbations and behavior under interactions and disorder
\citep{mahan2000many,fetter2012quantum}. In Fermi liquids, \textbf{$G(\omega,\mathbf{k})$}
has poles that directly correspond to the system's spectrum, with
the poles at $\omega=0$ defining the Fermi surface (FS). In contrast,
strongly correlated insulating phases like the Mott insulator have
\textbf{$G(\omega,\mathbf{k})$} that lacks poles but has zeros within
the energy gap. The locus of the zeros of \textbf{$G(\omega=0,\mathbf{k})$}
is similarly defined as the Luttinger surface (LS) \citep{dzyaloshinskii2003some}.
Since LSs correspond to the complete absence of quasiparticles, they
have traditionally been deemed as peculiar analytic features
of \textbf{$G(\omega,\mathbf{k}\mathbf{)}$} indicative of strong
correlations, but with little further fundamental or practical value
as they are unobservable and do not contribute to physical properties.

Recent breakthroughs, however, have dispelled this notion. Fabrizio
proposed a new class of quasiparticles
near LSs in strongly correlated systems that produce similar thermodynamic
properties as the quasiparticles in a Fermi liquid, such as the linear-in-$T$
specific heat \citep{fabrizio2022emergent}. This was followed by several seminal results within
the realm of interacting topological condensed matter. Wagner et al. showed that
Green's function zeros characterize the topology of Mott insulators
analogous to how poles characterize band insulators \citep{wagner2023mott}. Jinchao et al. \citep{zhao2023failure} and Bollmann et al. \cite{bollmann2024topological} discovered that the zeros of the single-particle Green's function modify the topological invariant, but the latter showed that the zeros do not affect the quantized topological response.
Blason and Fabrizio unified the role of poles and zeros in the topological
characterization of generic interacting insulators \citep{blason2023unified}. 

In contrast to these works, one of us showed that the boundary of a generic Weyl semimetal (WSM) -- a non-interacting topological phase --
hosts Luttinger arcs in addition to the well-known Fermi arcs \citep{obakpolor2022surface}. Nonetheless,
both structures connect surface projections
of Weyl nodes of opposite chirality. All these results place FSs and
LSs on similar footing, and are in similar spirit as
a pioneering result by Volovik that the same winding number
stabilizes FSs and LSs \citep{volovik2003universe}.

\begin{figure}[h]
\includegraphics[width=1.\columnwidth,height=1.\textheight,keepaspectratio]{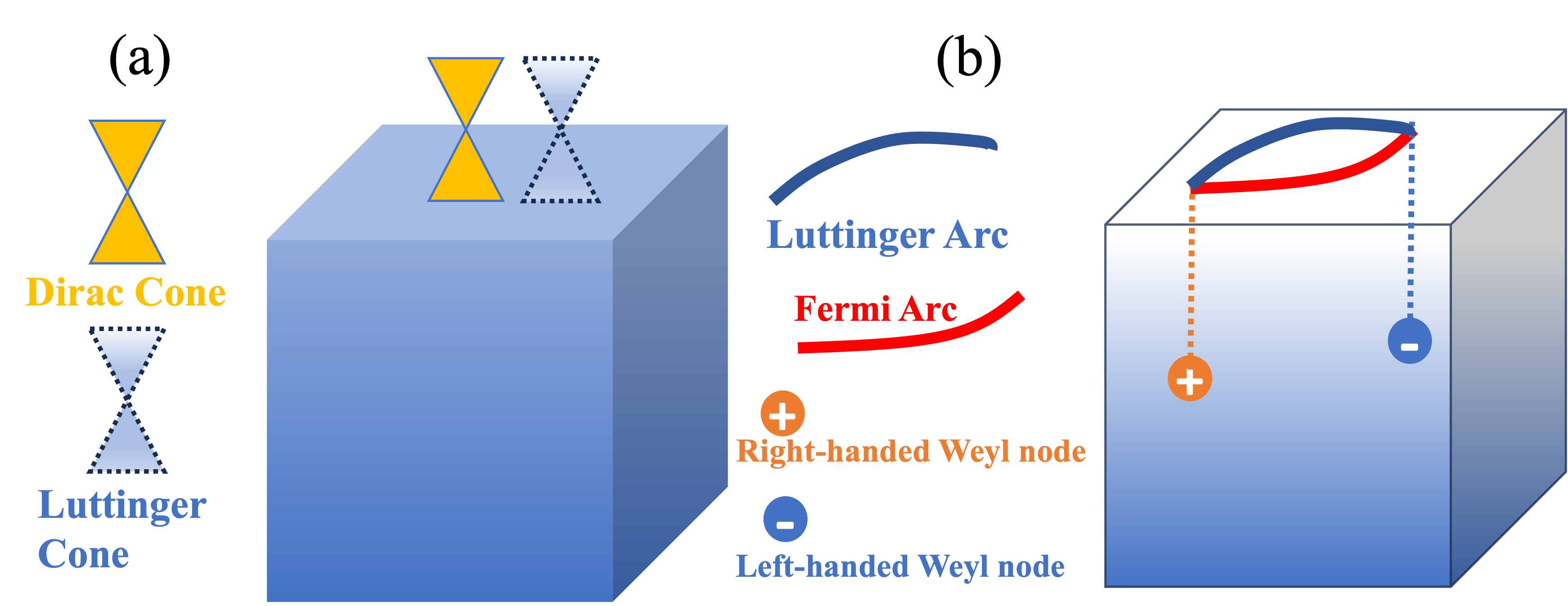}
\caption{Schematic of the nontrivial surface structures of a TI and a WSM. (a) Co-existing Dirac and Luttinger cones on a TI surface. (b) Co-existing Fermi and Luttinger arcs on the surface of a WSM, both connecting the surface projections of Weyl nodes of opposite chirality.}
\label{Fig.0}
\end{figure}

{\green{In this work, we generalize previous results on Luttinger arcs in Weyl semimetals to general non-interacting topological band structures, thus proving that boundary LSs are ubiquitous in non-interacting topological phases and interactions are not necessary for their existence.}} We begin by proving the existence of boundary LSs that respect the same bulk-boundary correspondence as the topological surface states {(}Fig. \ref{Fig.0}{)}. As a result,
a bulk topological phase would trigger a reconstruction or "Lifshitz transition" of the
LSs. We then establish two profound consequences of this behavior. First, we show that boundary LSs, remarkably, survive symmetry-breaking perturbations on the surface that destroy the topological surface states. This behavior starkly contradicts the
prevailing notion that single-particle boundary properties are oblivious to the bulk topology once the surface states have been gapped out. On the other hand, it is similar in spirit, for instance, to the discontinuity in the surface Hall conductivity of a 3D topological insulator (TI) that is tuned across a bulk topological transition with the surface Dirac node gapped out throughout the process \citep{hasan2010colloquium,qi2011topological}. Second, we show that the LSs reveal a disagreement between two routes for obtaining a low-energy surface theory in topological band structures, leading to unusual thermodynamic properties, such as a negative surface specific heat.
We prove all these properties using generic topological arguments and substantiate them numerically using lattice models of 3D time-reversal symmetric $\mathbb{Z}_2$ TIs and WSMs perturbed by surface magnetism and surface superconductivity, respectively. We close by concretizing the results in the context of Bi$_2$Te$_3$/MnBi$_2$Te$_4$ heterostructures.

\section{Robust boundary LSs}

We first show that topological band structures not only host boundary LSs that obey the same bulk-boundary correspondence as the surface states, generalizing the prediction of surface Luttinger arcs in WSMs \citep{obakpolor2022surface}, but that the LSs have a surprising robustness -- they survive even when surface perturbations gap out the surface states.

Consider non-interacting electrons in a $d$-dimensional system with a $(d-1)$-dimensional boundary.
We label an arbitrary set of  degrees of freedom near the boundary as the "surface" and refer to the rest as the "bulk." The full Hamiltonian can be decomposed as
\begin{equation}
\label{eq:H}
H(\bk) = \begin{pmatrix}
H_S(\bk) & h(\bk) \\
h^\dagger(\bk) & H_B(\bk)
\end{pmatrix}
\end{equation}
where $H_S(\bk)$ and $H_B(\bk)$ are Bloch Hamiltonians for the surface and the bulk, respectively, $h(\bk),h^\dagger(\bk)$ describe the coupling between them, and $\bk$ is the momentum in the surface Brillouin zone (sBZ). The full and bulk Green's functions at complex frequency $z$ are $G(z,\bk)=\left(z-H(\bk)\right)^{-1}$, $G_B(z,\bk)=\left(z-H_B(\bk)\right)^{-1}$, while the effective surface Green's function obtained by integrating out the bulk degrees of freedom is (see App. \ref{sec:sgf}):
\begin{equation}
\label{eq:sgf}
G_s(z,\bk) = \left[z-H_S(\bk)-h(\bk)G_B(z,\bk)h^\dagger(\bk)\right]^{-1}
\end{equation}

The LS is defined as the locus of zero eigenvalues of $G_s(0,\bk)$. To see it emerge, we first note that $G_s^{-1}(z,\bk)$ 
 obeys the condition: 
\begin{align}
\det G_{s}\left(z,\mathbf{k}\right) = \frac{\det G\left(z,\mathbf{k}\right)}{\det G_{B}\left(z,\mathbf{k}\right)},
\label{eq:DET}
\end{align}
Clearly, every pole of $G_B(z,\bk)$ causes $\det G_s\left(z,\mathbf{k}\right)$ to vanish provided $\det G\left(z,\mathbf{k}\right)$ stays finite, implying a zero eigenvalue of $G_s(z,\bk)$. Physically, this means surface FSs of $H_B(\bk)$ that disappear upon coupling $H_B(\bk)$ and $H_S(\bk)$ produce boundary LSs. At $\bk$ points where $G\left(z,\mathbf{k}\right)$ and $G_B\left(z,\mathbf{k}\right)$ have equal numbers of poles, the situation is subtler. These poles cancel in Eq. \ref{eq:DET} and mandate $\det G_s(z,\bk)$ to be finite and non-zero, hence forcing $G_s(z,\bk)$ to have equal numbers of diverging (pole) and vanishing (zero) eigenvalues, but does not guarantee the presence of either. On physical grounds, though, we expect surface poles of $G(z,\bk)$, i.e., poles whose corresponding eigenstates have finite weight on the surface in the thermodynamic limit, to be inherited by $G_s(z,\bk)$, since they represent physical surface FSs. We prove this explicitly in Appendix \ref{app:inheritance}. To ensure a finite $\det G_s(z,\bk)$, $G_s(z,\bk)$ must then also have an equal number of vanishing eigenvalues, thereby defining a LS that coincides with the surface FS. Thus, in all cases, the LS traces the surface FSs of $G_B(z,\bk)$ and therefore obeys the same bulk-boundary correspondence as the latter.

This construction instantly yields our first result: LSs persists even when surface FSs are destroyed by symmetry breaking perturbations on the surface. Such perturbations gap out the boundary states of the full system, represented by poles of $G(0,\mathbf{k})$ whose eigenfunctions are localized at the boundary. However, since the bulk degrees of freedom remains unaffected, the poles of $G_B(0,\mathbf{k})$ persist, and so do the zeros of the $G_s(0,\mathbf{k})$. We demonstrate this on a lattice model that realizes 3D TIs and WSMs in next section.

\section{Application to TIs and WSMs}

Consider an orthorhombic lattice model defined by the 3D Bloch Hamiltonian \citep{giwa2021fermi}: 
\begin{equation}
H_0(\mathbf{k}^{3D})=\tau_{x}\boldsymbol{\sigma}\cdot\mathbf{d}\left(\mathbf{k}^{3D}\right)+\tau_{z}m(\mathbf{k}^{3D})-\theta\tau_{y}\sigma_{z}-\mu,\label{eq:hami0}
\end{equation}
where $m(\mathbf{k}^{3D})=m_{0}-\sum_{i\in x,y,z}\beta_{i}\cos k_{i}$, $d_{i}\left(\mathbf{k}^{3D}\right)=v_{i}\sin\left(k_{i}\right)$ and $\tau_{i}$ ($\sigma_{i}$) are Pauli matrices in
orbital (spin) space. The wavevector is defined as $\mathbf{k}^{3D}\equiv \left(k_x, k_y, k_z\right)$. By tuning $\boldsymbol{\beta}$ and $\theta$, one can induce transitions between a trivial insulator, weak and strong TIs, and WSMs.

We first consider a strong TI, where we expect a 2D Dirac cone and a 2D Luttinger cone on the surface while time-reversal symmetry is preserved. If the symmetry is broken on the surface, we expect the Dirac cone to disappear but the Luttinger cone to survive. To uncover LSs, we calculate the surface Green's function $G_{s}\left(i \eta+E,k_x,k_y\right)$ \citep{sancho1985highly} recursively (see Appendix \ref{app:F}) and locate its zeros (poles) for a given $(k_x,k_y)$ by scanning for values of $E$ where the minimum (maximum) singular value of $G_{s}\left(i \eta+E,k_x,k_y\right)$ vanishes (diverges) as $\eta\to0^+$.

\begin{figure}[h]
\includegraphics[width=1.\columnwidth,height=1.\textheight,keepaspectratio]{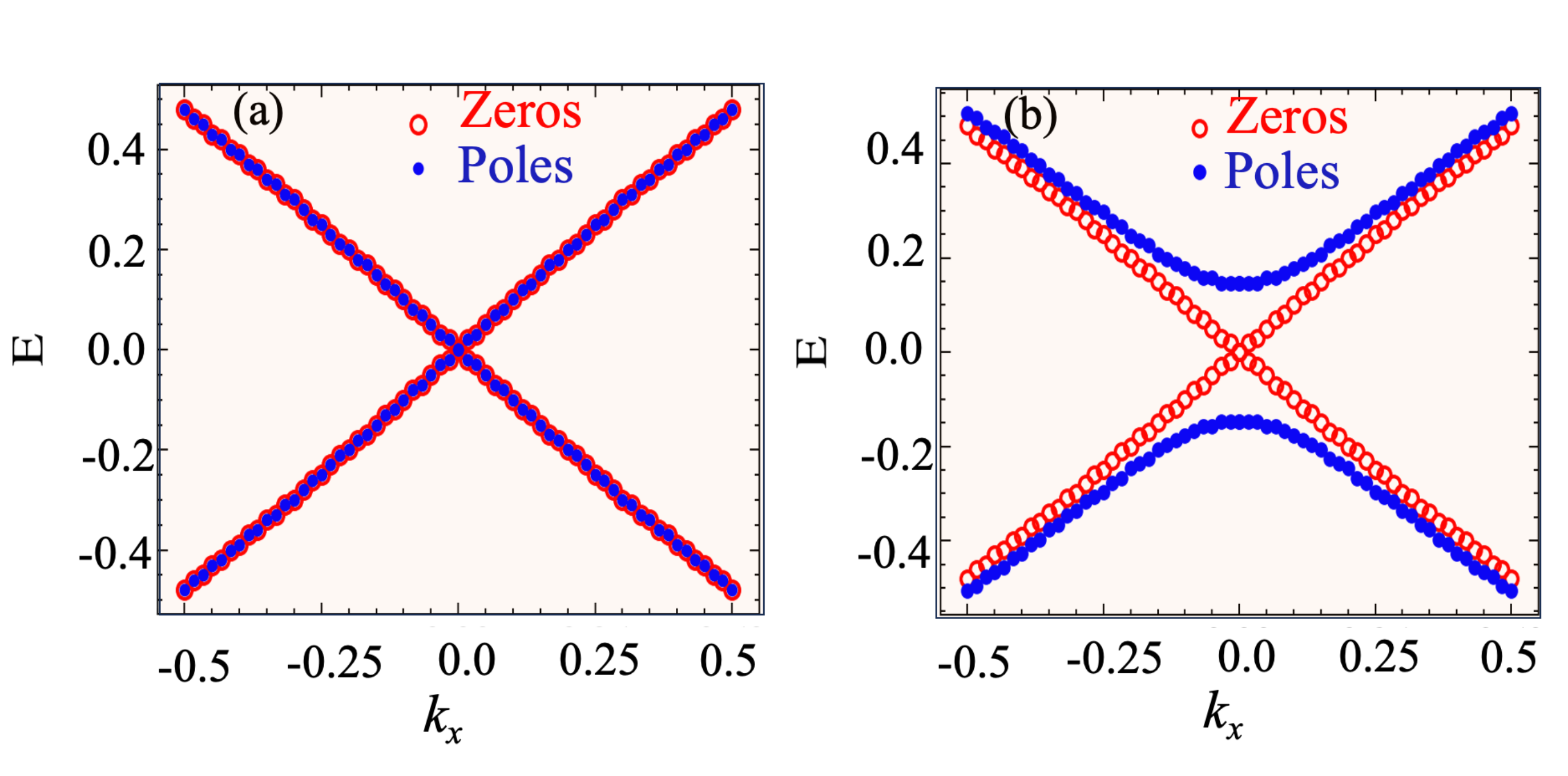} 
\caption{Green's function zeros and poles on the surface of TI without (a) and (b) with surface magnetization as a function of $k_{x}$ with $k_{y}=0$. The parameters
$m_{0}=1.8$, $\beta_{x,y,z}=1,1,0.5$, $v_{x,y,z}=1,1.5,1$, $\mu=0$,
$\theta=0$ and $m=0$ in (a) and $m=0.2$ in (b). Clearly, the magnetization destroys the Dirac cone (poles) but does not affect the Luttinger cone (zeros).}
\label{Fig.1} 
\end{figure}

The results are illustrated in Fig. \ref{Fig.1} (a) for $k_y=0$ and agree precisely with this prediction. Clearly, both the poles and zeros form
a cone-like dispersion in the sBZ in the absence of the symmetry-breaking surface perturbation.
To break the symmetry on the surface, we introduce a surface magnetization $H_{BTR}\equiv m\sigma_{z}$, which changes the surface Green's function to $\tilde{G}_{s}=\left(G_{s}^{-1}+H_{BTR}\right)^{-1}$. As shown in Fig.~\ref{Fig.1}(b), the Dirac cone, formed by the poles of $G_{s}$, is immediately gapped, but the Luttinger cone, arising from the zeros
of $G_{s}$, persists. In Appendix \ref{app:B}, we further confirm the persistence of the Luttinger cone by calculating the Volovik winding number that endows FSs and LSs with topological stability \citep{volovik2003universe}. 

As a second test case, we consider WSMs, distinguished
by their gapless bulk with chiral topological charges known as Weyl nodes, and surface Fermi arcs connecting the projections of Weyl nodes of opposite chiralities onto the sBZ \citep{hosur2013recent,armitage2018weyl}. The Fermi arcs can also be understood as collections of Bloch momenta
on the sBZ where $G_{s}\left(z=0,k_x,k_y\right)$ has poles. Similarly,
the surfaces of WSMs host Luttinger arcs, defined as
regions of the sBZ where $G_{s}\left(z=0,k_x,k_y\right)$ has zeros, which cause the non-interacting surface electrons to effectively acquire strong interactions mediated by bulk states \citep{obakpolor2022surface}. In contrast to the Dirac point on a TI surface, the Fermi arcs cannot be gapped by a simple translationally invariant magnetization perturbation. However, if the WSM is time-reversal symmetric, the Fermi arcs can be gapped by conventional superconductivity, which gives us an opportunity to inspect LSs under another physically relevant perturbation.

\begin{figure}[h]
\includegraphics[width=1\columnwidth,height=1.2\textheight,keepaspectratio]{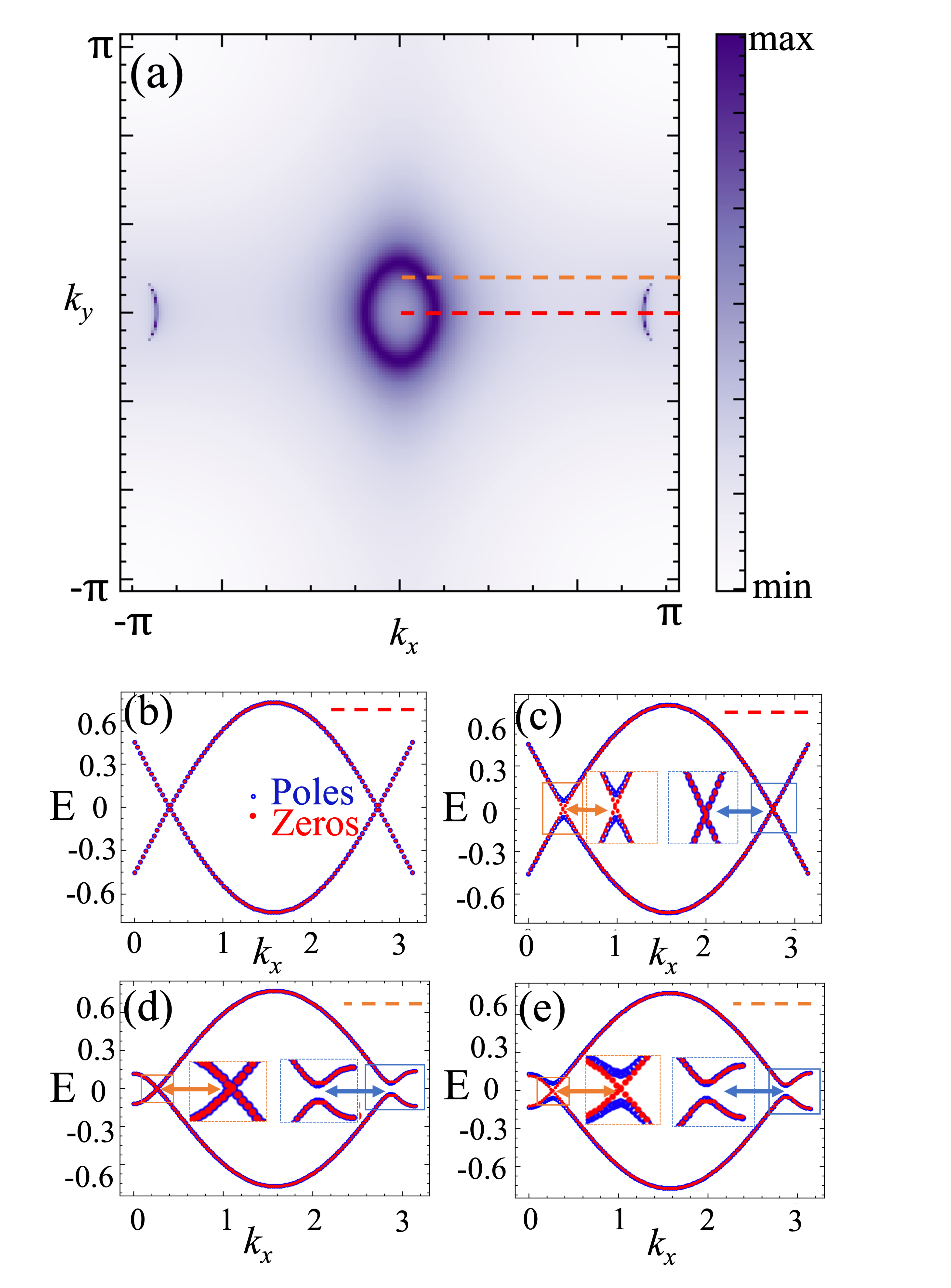}
\caption{ Zeros and poles of the surface Green's function in WSMs with and without a surface gap induced by superconductivity, along with Fermi arcs in WSMs. (a) Shows Fermi arcs on the sBZ of the WSM. (b) Depicts the distribution of Green's function zeros and poles as a function of $k_{x}\in\left[0,\pi\right]$ with $k_{y}=0$ (red dashed line in (a)) in the absence of surface superconductivity, $\delta=0$. Other  parameters are $m_{0}=3.3$, $\beta_{x,y,z}=0.856,1.178,3$, $v_{x,y,z}=1.18,0.856,1$, $\mu=0$, and $\theta=0.452$. (c) Same distribution as in (b), but with surface superconductivity, $\delta=0.1$. The distribution of Green's function zeros and poles is symmetric around the axis $k_{x}=0$. (d,e) Same distribution as in (b,c), respectively, but with $k_{y}=0.4$, as denoted by the orange dashed line in (a). The insets in (c-e) provide zoomed-in views of the boxed regions.}
\label{Fig.2} 
\end{figure}

Similarly to the procedure for the TI,
we recursively calculate surface Green's function for the lattice model in the WSM phase. Fermi arcs emerge within the sBZ as illustrated
in Fig.\ref{Fig.2}(a), while Luttinger arcs defined by Green's function zeros appear on top of the Fermi arcs, as shown in Fig.\ref{Fig.2} (b,d) for two values of $k_y$.
In the presence of s-wave surface superconductivity, either induced by proximity to another superconductor or intrinsically, as examined in recent works \cite{huang2019proximity,croitoru2020microscopic,yi2024interface,kuibarov2024evidence}, we can write the Bogoliubov-de Gennes (BdG) Hamiltonian as:
\begin{equation}
H_{\text{BdG}}(\mathbf{k}^{3D})=\begin{pmatrix}H_{\text{W}}(\mathbf{k}^{3D}) & \Delta\\
\Delta^{\dagger} & -H_{\text{W}}^{*}(-\mathbf{k}^{3D})
\end{pmatrix},\label{eq:hbdg}
\end{equation}
where the Hamiltonian $H_{\text{W}}(\mathbf{k}^{3D})$ is defined in the
same manner as in Eq. (\ref{eq:hami0}), with parameters adjusted
to represent the model in the WSM phase. Here, $\Delta=\delta s_{y}\sigma_{y}$
with $s_{y}$ acting on Nambu basis represents the superconducting pairing
potential, which only has a non-zero value on the surface of the lattice
model. As depicted in Fig.\ref{Fig.2} (c,e), the Fermi arcs near $\left(k_{x},k_{y}\right)=\left(0,0\right)$
are gapped. However, the Luttinger arcs persist as anticipated.

\section{Surface vs low-energy theory}

Consider a gapped non-interacting topological phase on a slab. Conventionally, the gapped, plane waves are referred to as bulk states and the gapless, evanescent waves localized near the surface are called surface states. Besides being parametrically separated in energy, the two classes of eigenstates have vanishing overlap in the thermodynamic limit. Thus, na\"{i}vely, integrating out either high energy states and low energy states of the far surface, if any, or integrating out bulk degrees of freedom should produce identical descriptions of the near-surface physics in the thermodynamic limit. This logic also applies to gapless systems such as WSMs if one avoids projections of the bulk Weyl nodes in the sBZ. 

We now prove this expectation incorrect. In particular, we show that the two approaches yield distinct values for thermodynamic properties, with the mismatch coming from LSs. Thus, the surface and bulk degrees of freedom are intimately coupled even though energy eigenstates traditionally associated with the surface and the bulk are separated in real space and in energy. 

\begin{figure}[h]
\includegraphics[width=1\columnwidth,height=1.2\textheight,keepaspectratio]{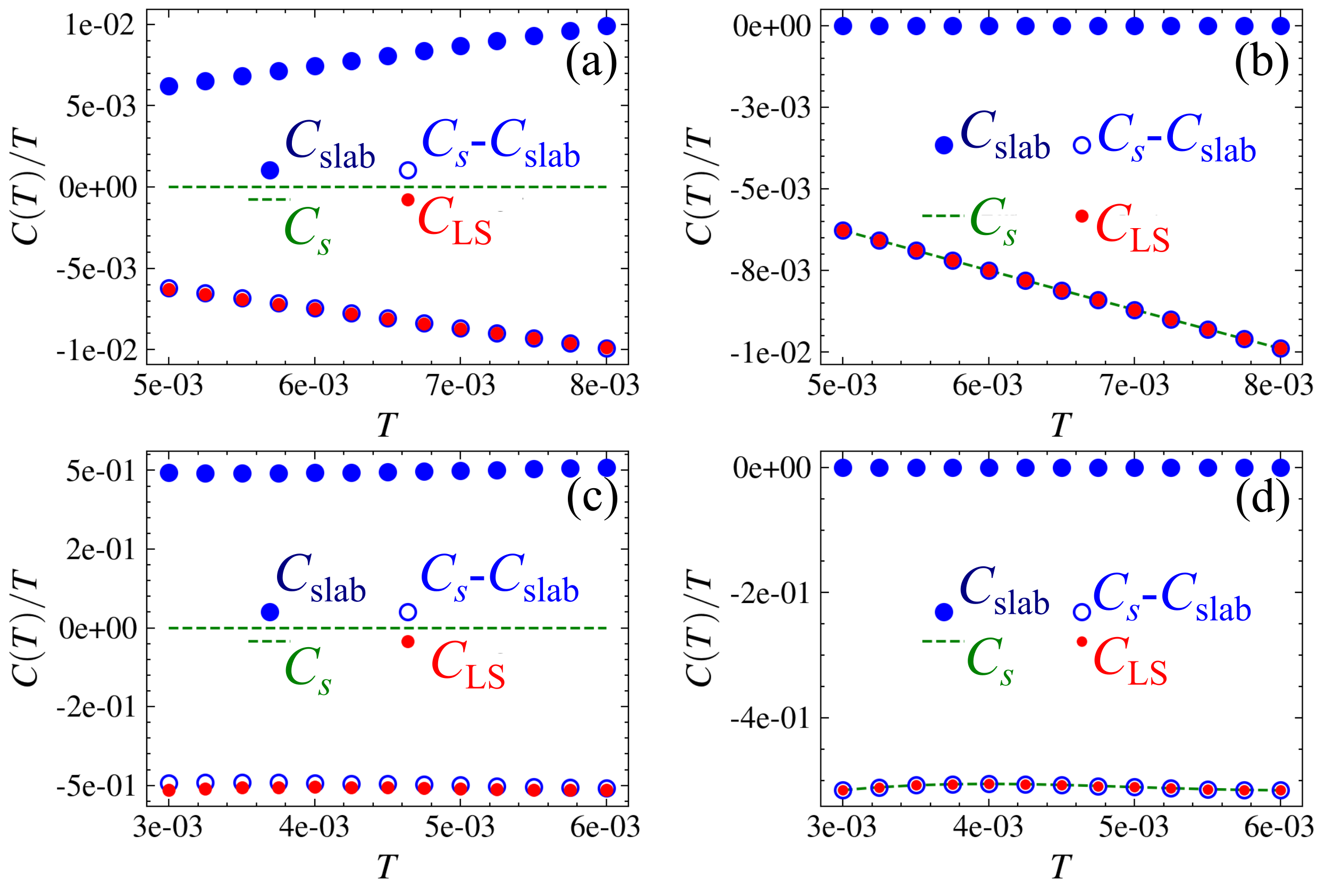}
\caption{Heat capacity $C\left(T\right)$ divided by temperature $T$ as a function of $T$. Panels (a) and (b) present results for the TI with and without surface perturbation, with all parameters as in Fig. \ref{Fig.1}. Panels (c) and (d) show results for the WS under the same conditions, with all parameters as in Fig. \ref{Fig.2}. $C_{\text{slab}}$, $C_s$, $C_s-C_{\text{slab}}$, and $C_{\text{LS}}$ represent the results from slab geometry with 40 layers in the $z$-direction, the effective surface Green's function, the difference between these two, and the Luttinger surface, respectively.}
\label{fig:specific-heat} 
\end{figure}

To see this mismatch explicitly, consider the single-particle free energy from a generic Green's function $\mathcal{G}(z,\bk)$, $\mathcal{F}=T\sum_{i\omega_{n}}\intop_{\mathbf{k}}\ln\det \mathcal{G}(i\omega_{n},\bk)$. Writing $\det \mathcal G(z,\bk)$ in terms of its poles $P_{n}(\mathbf{k})$ and zeros $Z_{m}(\mathbf{k})$, $\det \mathcal G(z,\bk)=\frac{\prod_{m}[z-Z_{i}(\mathbf{k})]}{\prod_{j}[z-P_{n}(\mathbf{k})]}$, gives the specific heat: 
\begin{align}
\mathcal C\left(T\right) = -T\frac{\partial^{2}\mathcal F}{\partial T^{2}} &=\intop_{\mathbf{k}}\sum_{n}\left[\frac{P_{n}(\mathbf{k})}{2T}\text{sech}\frac{P_{n}(\mathbf{k})}{2T}\right]^{2}\nonumber \\
&\quad-\sum_{m}\left[\frac{Z_{m}(\mathbf{k})}{2T}\text{sech}\frac{Z_{m}(\mathbf{k})}{2T}\right]^{2}.
\label{eq:heatc}
\end{align}
Clearly, only poles and zeros within $\sim T$ contribute appreciably to $\mathcal C(T)$. Evaluating $\mathcal C(T)$ at low $T$ using the full ($G$), bulk ($G_B$) and effective surface ($G_s$) Green's functions unveils the mismatch. In particular, for each $\bk$ point where the bulk plane waves and far-surface states are gapped with a gap $\gg T$ and the near-surface is gapless, setting $\mathcal G = G$ yields the specific heat from the near-surface states, $\mathcal C = C_\text{pl} + C_\text{far} + C_\text{near} \approx C_\text{near}$ as $C_\text{pl}$ and $C_\text{far}$ are negligible. In contrast, choosing $\mathcal G=G_s$ includes a negative contribution from the zeros or the LSs: $\mathcal C = C_s = C_\text{near} + C_{\text{LS}}$, where $C_{\text{LS}} < 0$. In fact, if the near-surface states are gapped out by a symmetry-breaking surface perturbation, then $C_\text{near}/T\to0$ as $T\to0$ while $C_s<0$. Physically, the bizarre negative specific heat represents the \textit{reduction} in total specific heat when the surface FSs of $H_B(\bk)$ are gapped out upon coupling the new surface layer $H_S$ to $H_B$ via $h,h^\dagger$. Alternately, it can be viewed as the specific heat due to the appearance of the LS. We demonstrate these points numerically using lattice models for a TI and WSM in Fig. \ref{fig:specific-heat}; for the latter, we separately calculate the specific heat of the gapless bulk states by applying periodic boundary conditions and subtract it from the total to isolate the Fermi arc contributions.
As shown in Figs. \ref{fig:specific-heat}(a,c), $C_\text{slab}/T = (C_\text{pl} + C_\text{far} + C_\text{near})/T>0$ for the TI slab and WSM slab without the near-surface perturbation, while $C_s/T=0$. After introducing surface perturbation, $C_\text{slab}/T\approx0$ for both the TI and WSM slabs and $C_s/T<0$, as shown in Figs. \ref{fig:specific-heat}(b) and \ref{fig:specific-heat}(d). In all cases, the difference, $(C_s-C_\text{slab})/T$, matches the LS contribution. 

{\green{While the reduction in heat capacities can also be explained without considering LSs, interpreting it through LSs highlights the fundamental distinction between surface theories and low-energy bulk theories in topological band structures. Identifying this discrepancy is not only conceptually interesting but also crucial for developing an accurate surface theory to study the surface properties of topological materials. This is especially true for topological semimetals, where the surface properties remain far less studied compared to the bulk due to the surface Hamiltonian being ill-defined. However, surface Green's functions, which help define LSs in this work, have yielded new insights on surface transport \cite{nomani2023intrinsic}  and superconductivity \cite{pal2022anomalous} in Weyl semimetals.}}

\section{Application to $\text{Bi}_2\text{Te}_3/ \text{MnBi}_2\text{Te}_4$}

To effectively "detect" LSs in experiment, we propose examining the specific heat difference between a TI, $\text{Bi}_2\text{Te}_3$ \cite{zhang2009topological}, and the same TI with a thin film of $\text{MnBi}_2\text{Te}_4$ \cite{chen2009experimental,li2019intrinsic,deng2020quantum,he2020mnbi2te4} deposited on top. Both $\text{Bi}_2\text{Te}_3$ and $\text{MnBi}_2\text{Te}_4$ are van der Waals coupled layered materials; given the recent advancements in fabricating van der Waals heterostructures, depositing the latter on top of the former is conceivable. The thin film should ideally have an even number of layers; an odd-layer magnetic $\text{MnBi}_2\text{Te}_4$ is a Chern insulator with edge states that also yield a contribution to the specific heat which must be subtracted off separately. In addition, the bare TI $\text{Bi}_2\text{Te}_3$ should be relatively thin so that the surface contribution to the specific heat is distinguishable from the bulk contribution -- a nontrivial requirement since bare $\text{Bi}_2\text{Te}_3$ has bulk Fermi pockets. When $\text{MnBi}_2\text{Te}_4$ is in its magnetized phase, the surface states of $\text{Bi}_2\text{Te}_3$ are expected to be gapped out, leading to a reduction in the specific heat. This reduction can alternately be interpreted as the negative specific heat due to the LSs that develop in the $\text{MnBi}_2\text{Te}_4$ thin film upon coupling to the TI. In this setup, the $\text{MnBi}_2\text{Te}_4$ thin film functions as the surface in the context of this work.

\section{Summary} 

We demonstrated that
in noninteracting topological phases of matter, LSs, formed by vanishing eigenvalues of the surface Green's function, coexist with topological surface states. 
For instance, the surface of a TI hosts both a Dirac cone and a Luttinger cone, while a WSM surface accommodates Fermi and Luttinger arcs. Their coexistence is mandated by no-go theorems that the surface Green's functions must obey but surface Hamiltonians evade in topological phases. Remarkably, surface perturbations that break the symmetries protecting the topological phase are unable to destroy the Luttinger surfaces even if they gap out the surface Fermi surfaces. {\blue{We then show that boundary LSs are absent in the low-energy theory of topological phases derived by integrating out higher-energy bulk states, but occur in the surface theory through the integration of bulk degrees of freedom. The existence of boundary LSs is tightly connected to the bulk topology; thus, the absence of LSs in the low-energy theory of topological phases demonstrates that the surface theory is different from the low-energy theory, although they are naively considered equivalent. Our findings shed new light on the role of Green's function zeros in topological condensed matter systems.}}

\section{acknowledgments}

 The authors are grateful to Steven Kivelson and especially Hridis Pal for fruitful discussions. K.C. and P.H. acknowledge support from National Science Foundation grant no. DMR 2047193. 

\bibliographystyle{unsrt}
\bibliography{Lutlib}

\begin{thebibliography}{10}

\bibitem{hasan2010colloquium}
M~Zahid Hasan and Charles~L Kane.
\newblock Colloquium: topological insulators.
\newblock {\em Reviews of modern physics}, 82(4):3045, 2010.

\bibitem{qi2011topological}
Xiao-Liang Qi and Shou-Cheng Zhang.
\newblock Topological insulators and superconductors.
\newblock {\em Reviews of Modern Physics}, 83(4):1057, 2011.

\bibitem{volovik2003universe}
Grigory~E Volovik.
\newblock {\em The universe in a helium droplet}, volume 117.
\newblock OUP Oxford, 2003.

\bibitem{armitage2018weyl}
N.~P. Armitage, E.~J. Mele, and Ashvin Vishwanath.
\newblock Weyl and dirac semimetals in three-dimensional solids.
\newblock {\em Rev. Mod. Phys.}, 90:015001, Jan 2018.

\bibitem{graf2013bulk}
Gian~Michele Graf and Marcello Porta.
\newblock Bulk-edge correspondence for two-dimensional topological insulators.
\newblock {\em Communications in Mathematical Physics}, 324:851--895, 2013.

\bibitem{rhim2018unified}
Jun-Won Rhim, Jens~H Bardarson, and Robert-Jan Slager.
\newblock Unified bulk-boundary correspondence for band insulators.
\newblock {\em Physical Review B}, 97(11):115143, 2018.

\bibitem{volovik2017topological}
GE~Volovik.
\newblock Topological lifshitz transitions.
\newblock {\em Low Temperature Physics}, 43(1):47--55, 2017.

\bibitem{mahan2000many}
Gerald~D Mahan.
\newblock {\em Many-particle physics}.
\newblock Springer Science \& Business Media, 2000.

\bibitem{fetter2012quantum}
Alexander~L Fetter and John~Dirk Walecka.
\newblock {\em Quantum theory of many-particle systems}.
\newblock Courier Corporation, 2012.

\bibitem{dzyaloshinskii2003some}
Igor Dzyaloshinskii.
\newblock Some consequences of the luttinger theorem: The luttinger surfaces in
  non-fermi liquids and mott insulators.
\newblock {\em Physical Review B}, 68(8):085113, 2003.

\bibitem{fabrizio2022emergent}
Michele Fabrizio.
\newblock Emergent quasiparticles at luttinger surfaces.
\newblock {\em Nature communications}, 13(1):1561, 2022.

\bibitem{wagner2023mott}
Niklas Wagner, Lorenzo Crippa, Adriano Amaricci, Philipp Hansmann, Marcel
  Klett, EJ~K{\"o}nig, Thomas Sch{\"a}fer, D~Di Sante, Jennifer Cano,
  AJ~Millis, et~al.
\newblock Mott insulators with boundary zeros.
\newblock {\em Nature Communications}, 14(1):7531, 2023.

\bibitem{zhao2023failure}
Jinchao Zhao, Peizhi Mai, Barry Bradlyn, and Philip Phillips.
\newblock Failure of topological invariants in strongly correlated matter.
\newblock {\em Phys. Rev. Lett.}, 131:106601, Sep 2023.

\bibitem{bollmann2024topological}
Steffen Bollmann, Chandan Setty, Urban~FP Seifert, and Elio~J K{\"o}nig.
\newblock Topological green's function zeros in an exactly solved model and
  beyond.
\newblock {\em Physical Review Letters}, 133(13):136504, 2024.

\bibitem{blason2023unified}
Andrea Blason and Michele Fabrizio.
\newblock Unified role of green's function poles and zeros in correlated
  topological insulators.
\newblock {\em Physical Review B}, 108(12):125115, 2023.

\bibitem{obakpolor2022surface}
Osakpolor~Eki Obakpolor and Pavan Hosur.
\newblock Surface luttinger arcs in weyl semimetals.
\newblock {\em Physical Review B}, 106(8):L081112, 2022.

\bibitem{giwa2021fermi}
Rauf Giwa and Pavan Hosur.
\newblock Fermi arc criterion for surface majorana modes in superconducting
  time-reversal symmetric weyl semimetals.
\newblock {\em Physical Review Letters}, 127(18):187002, 2021.

\bibitem{sancho1985highly}
MP~Lopez Sancho, JM~Lopez Sancho, JM~Lopez Sancho, and J~Rubio.
\newblock Highly convergent schemes for the calculation of bulk and surface
  green functions.
\newblock {\em Journal of Physics F: Metal Physics}, 15(4):851, 1985.

\bibitem{hosur2013recent}
Pavan Hosur and Xiaoliang Qi.
\newblock Recent developments in transport phenomena in weyl semimetals.
\newblock {\em C. R. Phys.}, 14(9):857--870, 2013.
\newblock Topological insulators / Isolants topologiques.

\bibitem{huang2019proximity}
Ce~Huang, Benjamin~T Zhou, Huiqin Zhang, Bingjia Yang, Ran Liu, Hanwen Wang,
  Yimin Wan, Ke~Huang, Zhiming Liao, Enze Zhang, et~al.
\newblock Proximity-induced surface superconductivity in dirac semimetal
  cd3as2.
\newblock {\em Nature communications}, 10(1):2217, 2019.

\bibitem{croitoru2020microscopic}
Mikhail~D Croitoru, AA~Shanenko, Y~Chen, Alexei Vagov, and J~Albino Aguiar.
\newblock Microscopic description of surface superconductivity.
\newblock {\em Physical Review B}, 102(5):054513, 2020.

\bibitem{yi2024interface}
Hemian Yi, Yi-Fan Zhao, Ying-Ting Chan, Jiaqi Cai, Ruobing Mei, Xianxin Wu,
  Zi-Jie Yan, Ling-Jie Zhou, Ruoxi Zhang, Zihao Wang, et~al.
\newblock Interface-induced superconductivity in magnetic topological
  insulators.
\newblock {\em Science}, 383(6683):634--639, 2024.

\bibitem{kuibarov2024evidence}
Andrii Kuibarov, Oleksandr Suvorov, Riccardo Vocaturo, Alexander Fedorov, Rui
  Lou, Luise Merkwitz, Vladimir Voroshnin, Jorge~I Facio, Klaus Koepernik,
  Alexander Yaresko, et~al.
\newblock Evidence of superconducting fermi arcs.
\newblock {\em Nature}, 626(7998):294--299, 2024.

\bibitem{nomani2023intrinsic}
Aymen Nomani and Pavan Hosur.
\newblock Intrinsic surface superconducting instability in type-i weyl
  semimetals.
\newblock {\em Physical Review B}, 108(16):165144, 2023.

\bibitem{pal2022anomalous}
Hridis~K Pal, Osakpolor~Eki Obakpolor, and Pavan Hosur.
\newblock Anomalous surface conductivity of weyl semimetals.
\newblock {\em Physical Review B}, 106(24):245410, 2022.

\bibitem{zhang2009topological}
Haijun Zhang, Chao-Xing Liu, Xiao-Liang Qi, Xi~Dai, Zhong Fang, and Shou-Cheng
  Zhang.
\newblock Topological insulators in bi2se3, bi2te3 and sb2te3 with a single
  dirac cone on the surface.
\newblock {\em Nature physics}, 5(6):438--442, 2009.

\bibitem{chen2009experimental}
YL~Chen, James~G Analytis, J-H Chu, ZK~Liu, S-K Mo, Xiao-Liang Qi, HJ~Zhang,
  DH~Lu, Xi~Dai, Zhong Fang, et~al.
\newblock Experimental realization of a three-dimensional topological
  insulator, bi2te3.
\newblock {\em science}, 325(5937):178--181, 2009.

\bibitem{li2019intrinsic}
Jiaheng Li, Yang Li, Shiqiao Du, Zun Wang, Bing-Lin Gu, Shou-Cheng Zhang,
  Ke~He, Wenhui Duan, and Yong Xu.
\newblock Intrinsic magnetic topological insulators in van der waals layered
  mnbi2te4-family materials.
\newblock {\em Science Advances}, 5(6):eaaw5685, 2019.

\bibitem{deng2020quantum}
Yujun Deng, Yijun Yu, Meng~Zhu Shi, Zhongxun Guo, Zihan Xu, Jing Wang, Xian~Hui
  Chen, and Yuanbo Zhang.
\newblock Quantum anomalous hall effect in intrinsic magnetic topological
  insulator mnbi2te4.
\newblock {\em Science}, 367(6480):895--900, 2020.

\bibitem{he2020mnbi2te4}
Ke~He.
\newblock Mnbi2te4-family intrinsic magnetic topological materials.
\newblock {\em npj Quantum Materials}, 5(1):90, 2020.

\bibitem{gurarie2011single}
Victor Gurarie.
\newblock Single-particle green's functions and interacting topological
  insulators.
\newblock {\em Physical Review B}, 83(8):085426, 2011.

\bibitem{seki2017topological}
Kazuhiro Seki and Seiji Yunoki.
\newblock Topological interpretation of the luttinger theorem.
\newblock {\em Physical Review B}, 96(8):085124, 2017.

\bibitem{PhysRevResearch.5.033043}
Maximilian Kotz and Carsten Timm.
\newblock Topological classification of non-hermitian hamiltonians with
  frequency dependence.
\newblock {\em Phys. Rev. Res.}, 5:033043, Jul 2023.

\bibitem{zhou2019periodic}
Hengyun Zhou and Jong~Yeon Lee.
\newblock Periodic table for topological bands with non-hermitian symmetries.
\newblock {\em Physical Review B}, 99(23):235112, 2019.

\bibitem{neupane2016observation}
Madhab Neupane, Nasser Alidoust, M~Mofazzel Hosen, Jian-Xin Zhu, Klauss
  Dimitri, Su-Yang Xu, Nagendra Dhakal, Raman Sankar, Ilya Belopolski, Daniel~S
  Sanchez, et~al.
\newblock Observation of the spin-polarized surface state in a
  noncentrosymmetric superconductor bipd.
\newblock {\em Nature communications}, 7(1):13315, 2016.

\end{thebibliography}

\onecolumngrid

\section*{Appendix for Robust boundary LSs in topological band structures}

In Appendix A, we present two distinct methods for deriving an effective theory. Appendix B explains how the effective surface Green's function can inherit the poles of the full Green's function. In Appendix C, we show how the winding number confirms the presence of LSs on the surfaces of topological insulators. Appendix D explores the constraints imposed by time-reversal symmetry on the presence of poles and zeros in Green's functions in fermionic systems, extending to generalized time-reversal symmetry in non-Hermitian systems. Appendix E provides a concise review of the recursive Green's function technique used to calculate the surface Green's function in the main text. Finally, Appendix F illustrates how to reconstruct the surface Green's function from ARPES data.

\renewcommand{\thefigure}{S\arabic{figure}}
\setcounter{figure}{0} 

\appendix

\section{Surface theory vs low-energy}
\label{sec:sgf}
In this section, we contrast two approaches for deriving an effective theory -- by explicitly integrating out the bulk degrees of freedom, which yields the surface Green's function $G_s(z,\bk)$ in Eq. \ref{eq:sgf} as a certain Schur complement, and integrating out high-energy states.

\subsection{Integrating out bulk degrees of freedom}

We begin with the Hamiltonian a $L$-layer slab of a $d$-dimensional topological phase with $(d-1)$-dimensional surfaces:
\begin{equation}
H({\mathbf{k}})=\left(\begin{array}{cc}
H_S({\mathbf{k}}) & h({\mathbf{k}})\\
h^\dagger({\mathbf{k}}) & H_B({\mathbf{k}})
\end{array}\right).
\end{equation}
We will use $\bar{s},s$ to denote Grassman variables for fermionic degrees of freedom on the "surface" and $\bar{b},b$ to denote the same for all the other fermions that we will concisely refer to as the "bulk". Integrals will be written in shorthand
as $\intop_{\mathbf{k},i\omega_n}\equiv T\sum_{\omega_n=(2n+1)\pi T}\int\frac{d^{d-1}k}{(2\pi)^{d-1}}$.
In this notation, the Euclidean path integrals for the bulk and full systems are $Z_B=\int\mathcal{D}\left[\bar{b},b\right]\exp\left[-\mathcal{S}_{B}\left(\bar{b},b\right)\right]$
and $Z=\int\mathcal{D}\left[\bar{b},b,\bar{s},s\right]\exp\left[-\mathcal{S}\left(\bar{b},b,\bar{s},s\right)\right]$
where
\begin{align}
\mathcal{S}_B\left(\bar{b},b\right) & =\intop_{\mathbf{k},i\omega_n}\bar{b}_{\mathbf{k}}\left[i\omega_n-H_{B}(\bk)\right]b_{\mathbf{k}},\\
 & \equiv\intop_{\mathbf{k},i\omega_n}\bar{b}_{\mathbf{k}}\cdot\left[G_{B}(i\omega_n,\bk)\right]^{-1}\cdot b_{\mathbf{k}},\\
\mathcal{S}\left(\bar{b},b,\bar{s},s\right) & =\intop_{\mathbf{k},i\omega_n}\left(\bar{s}_{\mathbf{k}},\bar{b}_{\mathbf{k}}\right)\cdot\left[i\omega_n-H({\mathbf{k}})\right]\cdot\left(\begin{array}{c}
s_{\mathbf{k}}\\
b_{\mathbf{k}}
\end{array}\right),\\
 & \equiv\intop_{\mathbf{k},i\omega_n}\left(\bar{s}_{\mathbf{k}},\bar{b}_{\mathbf{k}}\right)\cdot\left[G(i\omega_n,\bk)\right]^{-1}\cdot\left(\begin{array}{c}
s_{\mathbf{k}}\\
b_{\mathbf{k}}
\end{array}\right).
\end{align}
Performing a standard Gaussian-Grassman integral over $b,\bar{b}$ in $Z$ yields an effective surface Green's function $G_s(i\omega_n,\bk)$:
\begin{align}
Z_s & =\frac{Z}{Z_{B}}\equiv\int\mathcal{D}\left[\bar{s},s\right]\exp\left[-\mathcal{S}_s\left(\bar{s},s\right)\right],\label{eq:Zs}\\
\mathcal{S}_s\left(\bar{s},s\right) & =\intop_{\mathbf{k},i\omega_n}\bar{s}_{\mathbf{k}}\cdot\left[i\omega_n-H_S(\bk)-h(\bk)G_B(i\omega_n,\bk)h(\bk)^{\dagger}\right]\cdot s_{\mathbf{k}},\label{eq:Ss}\\
\implies G_s(i\omega_n,\bk) & =\left[i\omega_n-H_S(\bk)-h(\bk)G_B(i\omega_n,\bk)h(\bk)^{\dagger}\right]^{-1}\label{eq:Gs}.
\end{align}
The above expression is rather standard and has the form of a Dyson equation for the surface degrees of freedom with self-energy $\Sigma(i\omega_n,\bk) = h(\bk)G_B(i\omega_n,\bk)h(\bk)^{\dagger}$.

Alternately, we can obtain $G_s(i\omega_n,\bk)$ by straightforward linear algebra. Performing standard block matrix inversion of $i\omega_n-H(\bk)$ in the basis $\left( \mid s(\bk)\rangle, \mid b(\bk)\rangle \right)^T$ yields (suppressing frequency and momentum labels for brevity),
\begin{equation}
G=\begin{pmatrix}
G_s & G_shG_B \\
G_Bh^\dagger G_s & G_B+G_Bh^\dagger G_s h G_B
\end{pmatrix}
\end{equation}
Clearly, $G_s(i\omega_n,\bk)$ is simply the sub-block of the $G(i\omega_n,\bk)$ corresponding to the surface degrees of freedom. In linear algebra parlance, $G_s^{-1}(i\omega_n,\bk)$ is known as the Schur complement of $G_B^{-1}(i\omega_n,\bk)$ in $G^{-1}(i\omega_n,\bk)$ and simply denoted
\begin{equation}
G_s^{-1}(i\omega_n,\bk) = G^{-1}(i\omega_n,\bk)/G_B^{-1}(i\omega_n,\bk)
\end{equation} 
In other words, integrating out $b$ yields the same effective surface Green's functions as projecting the full Green's function onto the $s$-states.

\subsection{Integrating out high-energy plane waves}

In the parlance of topological condensed matter, "bulk" and "surface", respectively, refer to plane waves traversing the bulk at high energy and evanescent waves localized at a surface with exponential decay into the bulk layers. Strictly speaking, in a slab geometry, the energy eigenstates are standing waves traversing the slab and superpositions of evanescent waves localized on opposite surfaces. However, in the thermodynamic limit, traveling plane waves and evanescent waves on each surface become energy eigenstates. We will work in this limit for simplicity.

Now, an alternate derivation of the effective surface theory by integrating out the plane waves is justified. Explicitly, we can spectrally decompose the total Hamiltonian as
\begin{equation}
H(\bk) = \sum_{j} \zeta_j^\text{near}(\bk){|\psi_j^\text{near}(\bk)\rangle\langle\psi_j^\text{near}(\bk)|} + \sum_k\zeta_k^\text{far}|\psi_k^\text{far}(\bk)\rangle\langle\psi_k^\text{far}(\bk)| + \sum_{i} \varepsilon_i(\bk){|\psi_i^\text{pl}(\bk)\rangle\langle\psi_i^\text{pl}(\bk)|}
\end{equation}
where $\zeta_j^\text{near}(\bk)$, $\zeta_j^\text{far}(\bk)$ and $\varepsilon_i(\bk)$ denote energies of the evanescent waves on the surface under consideration, evanescent waves on the opposite surface of the slab, and plane wave states, respectively. As the various sets of states are decoupled, integrating out the plane waves and far-surface evanescent waves is trivial and we are left with effective Hamiltonian and Green's function for the near surface:
\begin{align}
H^\text{near}(\bk) &= \sum_{j} \zeta_j(\bk){|\psi_j^\text{near}(\bk)\rangle\langle\psi_j^\text{near}(\bk)|} \\
G^\text{near}(i\omega_n,\bk) &= \sum_{j} \frac{|\psi_j^\text{near}(\bk)\rangle\langle\psi_j^\text{near}(\bk)|}{i\omega_n-\zeta_j(\bk)} \\
&\equiv \text{diag}\{\left(i\omega_n-\zeta_j(\bk)\right)^{-1};j=1\dots D\}
\end{align}
in the basis of $\psi_j^\text{near}(\bk)$, where $D$ is the number of degrees of freedom defined as the surface.

\section{Inheritance of surface poles by the surface Green's function}\label{app:inheritance}

In this section, we prove that the effective surface Green's function $G_s(z,\bk)$ studied in this paper inherits surface poles from the full Green's function $G(z,\bk)$, i.e., poles from the latter whose corresponding eigenfunctions have finite support on the surface degrees of freedom.

Similarly to $H(\bk)$, we can spectrally decompose $G(z,\bk)$ as
\begin{equation}
G(\bk) = \sum_{j} \frac{|\psi_j^\text{near}(\bk)\rangle\langle\psi_j^\text{near}(\bk)|}{z-\zeta_j^\text{near}(\bk)} + \sum_k\frac{|\psi_k^\text{far}(\bk)\rangle\langle\psi_k^\text{far}(\bk)|}{z-\zeta_k^\text{far}} + \sum_{i} \frac{|\psi_i^\text{pl}(\bk)\rangle\langle\psi_i^\text{pl}(\bk)|}{z-\varepsilon_i(\bk)}
\end{equation}

Now, the plane waves and far-surface evanesccent waves have a vanishing amplitude on the near surface in the thermodynamic limit. Thus, projecting onto the near surface gives
\begin{equation}
G_s(z,\bk) = \sum_{n,n'}\sum_j |s_n(\bk)\rangle \left( \frac{\langle s_n(\bk)|\psi_j^\text{near}(\bk)\rangle \langle\psi_j^\text{near}(\bk)|s_{n'}(\bk)\rangle}{z-\zeta_j(\bk)} \right) \langle s_{n'}(\bk)|
\end{equation}
where $|s_n(\bk)\rangle, n=1\dots D$ are basis states spanning the near surface. Taking the trace gives
\begin{equation}
\text{Tr}G_s(z,\bk) = \sum_j\frac{|\langle s_n(\bk)|\psi_j^\text{near}(\bk)\rangle|^2}{z-\zeta_j(\bk)}
\end{equation}
Clearly, every surface pole of $G(z,\bk)$, where $z-\zeta_j(\bk)$ vanishes with $|\langle s_n(\bk)|\psi_j^\text{ev}(\bk)\rangle|$ being finite, causes $\text{Tr}G_s(z,\bk)$ to diverge, indicating a pole in an eigenvalue of $G_s(z,\bk)$.

\section{Identifying the LSs through the topological winding number }
\label{app:B}

The stability of the FS is
derived from the topological property of the single-particle Green's
function $G(i\epsilon,\mathbf{k})$ with $\mathbf{k}$ and $i\epsilon$
denoting the Bloch momentum and imaginary frequency \citep{volovik2003universe}.
The corresponding topological invariant is a winding number defined
as follows: 
\begin{equation}
N_{W}=\int_{C}\frac{dl}{2i\pi}\text{tr}\left[G^{-1}\left(i\epsilon,\mathbf{k}\right)\partial_{l}G\left(i\epsilon,\mathbf{k}\right)\right],\label{eq:winding}
\end{equation}
where $C$ denotes a counterclockwise loop in the $\left(\epsilon,\mathbf{k}\right)$-space,
and $\text{tr}$ is the trace over spin, orbital, or other indices.
The winding number can also be expressed as $\int_{C}\frac{dl}{2i\pi}\partial_{l}\ln\text{det}G\left(i\epsilon,\mathbf{k}\right)$.
Considering a simple single-particle Green's function $G\left(i\epsilon,\mathbf{k}\right)=\frac{1}{i\epsilon-\mathbf{v}_{F}\cdot\left(\mathbf{k}-\mathbf{k}_{F}\right)}$
with Fermi velocity $\mathbf{v}_{F}$ and Fermi momentum $\mathbf{k}_{F}$,
the resulting winding number is expressed as $N_{W}=\text{sgn}(\mathbf{v}_{F})\text{Link}(C,FS)$,
where $\text{Link}(C,FS)$ is the linking number between the
loop $C$ and the FS.

On the other hand, when the single-particle Green's function's zeros
are zero, they define the LS, where the Green's function
can be approximated as $G\left(i\epsilon,\mathbf{k}\right)=i\epsilon-\mathbf{v}_{F}\cdot\left(\mathbf{k}-\mathbf{k}_{L}\right)$
with Luttinger momentum $\mathbf{k}_{L}$. The resulting winding number
is expressed as $N_{W}=-\text{sgn}(\mathbf{v}_{F})\text{Link}(C,LS)$,
where $\text{Link}(C,LS)$ represents the link number between the
loop $C$ and the LS.

In this section, we demonstrate the utility of the winding number
in identifying the presence of zeros in the surface Green's function.
For the model considered in the previous section, numerical calculations
reveal that the unperturbed surface Green's function exhibits overlapped
zeros and poles, as illustrated in Fig. \ref{Fig.1}(a). Consequently,
the winding number is zero due to the cancellation between zeros and
poles. Upon introducing a surface perturbation to eliminate the poles,
the winding number for a loop with an imbalanced count of oppositely
moving zeros attains a nonzero integer value. This value equals the
difference between the numbers of left-moving and right-moving zeros
within the loop. 

\begin{figure}[h]
\includegraphics[width=0.6\columnwidth,height=0.6\textheight,keepaspectratio]{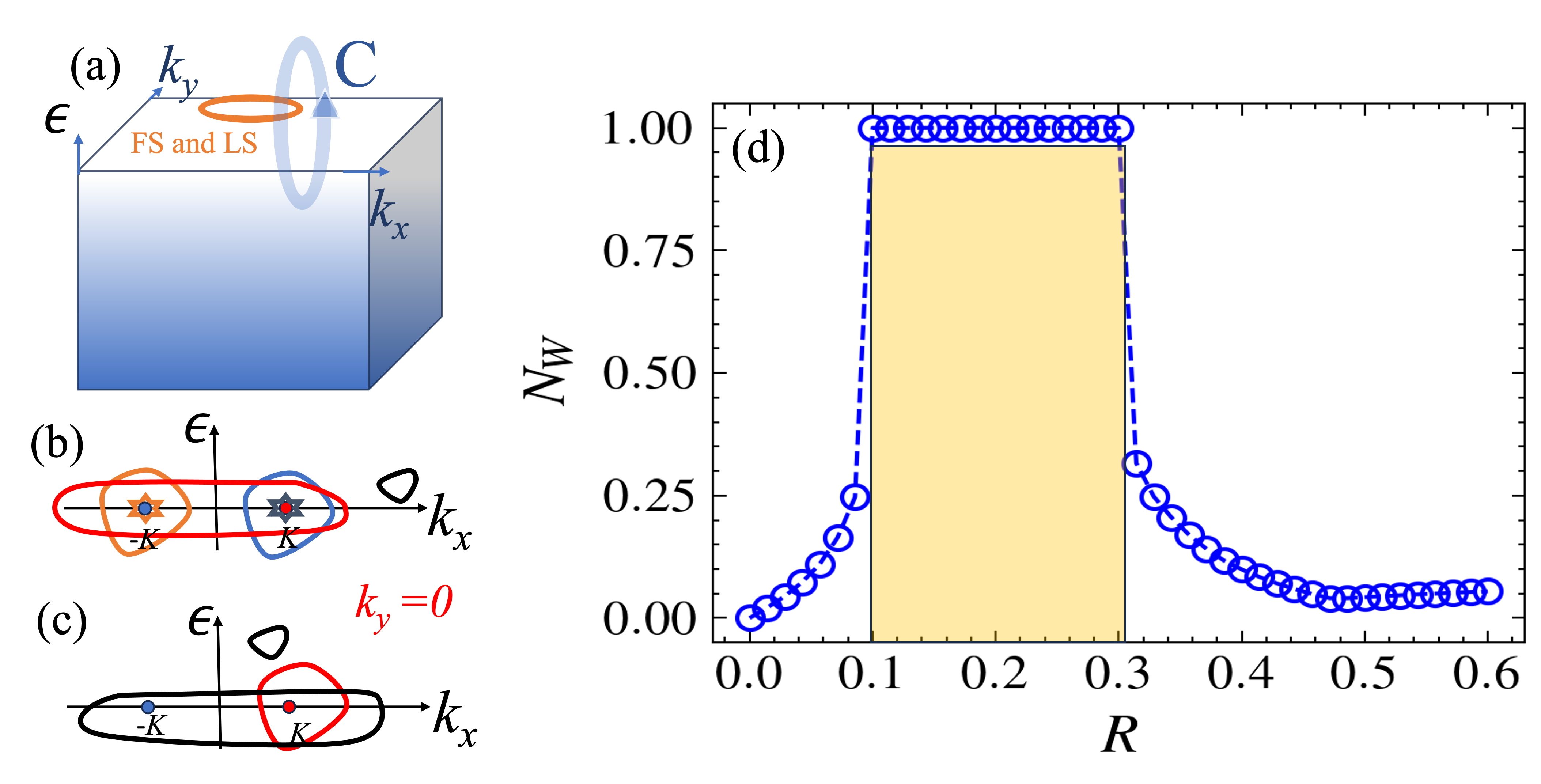} 
\caption{ Topological winding number as an indicator of the LS. (a-c) Schematic figures illustrating the FS
and LS and loops in the $(\mathbf{k},\epsilon)$-space.
The stars and points in (b) represent the Fermi points and Luttinger
points, respectively. The points in (c) also represent Luttinger points.
(d) The winding number as a function of the radius of the loop center
at $(k_{x},k_{y},\epsilon)=(0.2,0,0)$ with the poles gapped out. $N_W=1$ is a signature of the surviving LS. The parameters are $\mu=0.1$,
$m_{0}=1.8$, $\beta_{x,y,z}=1,1,0.5$, $v_{x,y,z}=1,1.5,1$, $\theta=0$,
and $m=0.5$.}
\label{Fig.4}
\end{figure}

To validate this assertion, we examine the surface Green's function
obtained from the lattice model in the topological insulating phase
with a nonzero chemical potential. In this scenario, the Dirac points
and Luttinger points evolve into Dirac circles and Luttinger circles,
respectively. Without loss of generality, we choose loops in the $(k_{x},\epsilon)$-plane.
For the unperturbed surface, there exists a right-moving ($sgn(\mathbf{v}_{F})>0$)
pole and a right-moving zero at the point $(K,0)$, as well as a left-moving
($sgn(\mathbf{v}_{F})<0$) pole and a left-moving zero at the point
$(-K,0)$\textbf{---}illustrated schematically in Fig. \ref{Fig.4}
(b). Consequently, the winding number is zero for any loop in the
$(k_{x},\epsilon)$-plane in this case.

For the perturbed surface, the poles are gapped out, but the right-moving
and left-moving zeros persist at the same points (Fig. \ref{Fig.4}
(c)). We focus on a loop characterized by points $(k_{x},\epsilon)=(k_{0}+R\cos\theta,R\sin\theta)$
with $\theta\in[-\pi,\pi)$. The winding number is illustrated in
Fig. \ref{Fig.4} (d), showing $N_{W}=1$ for the radius $R\in(k_{0}-K,k_{0}+K)$
(within the yellow rectangle). This range includes only a single right-moving
zero inside the loop. The winding number becomes zero for $0\leq R\leq k_{0}-K$
since no zeros are included inside the loop, and for $k_{0}+K\leq R$
because both a left-moving and a right-moving zero are included inside
the loop. In our numerical calculation, we use the parameter values
$k_{0}=2$ and $\pm K\approx\pm0.11$.

\section{Restrictions on zeros and poles by Time reversal symmetry }
\label{app:D}

For a single-particle Green's function, its determinant satisfies
the following general equation \citep{gurarie2011single,seki2017topological}:
\begin{equation}
\det G\left(z,\mathbf{k}\right)=\frac{\prod_{j}(z-Z_{j}\left(\mathbf{k}\right))}{\prod_{i}(z-P_{i}\left(\mathbf{k}\right))},\label{eq:detG}
\end{equation}
where $z$ represents a complex number, while $Z_{j}\left(\mathbf{k}\right)$
and $P_{i}\left(\mathbf{k}\right)$ be the zero and pole of a single-particle
Green's function at Bloch momentum $\mathbf{k}$, respectively. It's
important to note that $Z_{j}\left(\mathbf{k}\right)$ and $P_{i}\left(\mathbf{k}\right)$
are real-valued quantities. We then define the set $\mathfrak{Z}_{C}^{\lessgtr}=\{\mathbf{k}\in C\mid\hat{n}_{C}\cdot\nabla_{\mathbf{k}}Z\left(\mathbf{k}\right)\lessgtr0,Z\left(\mathbf{k}\right)=0\}$
and $\mathfrak{P}_{C}^{\lessgtr}=\{\mathbf{k}\in\text{C}\mid\hat{n}_{C}\cdot\nabla_{\mathbf{k}}P\left(\mathbf{k}\right)\lessgtr0,P\left(\mathbf{k}\right)=0\}$,
where $\hat{n}_{C}$ is unit direction vector along the directed loop,
we adopt a counterclockwise direction as the loop orientation throughout
this work, without loss of generality. We categorize the points in
set $\mathfrak{Z}_{C}^{<}$ as left-moving zeros, and those in set
$\mathfrak{Z}_{C}^{>}$ as right-moving zeros. Similarly, points in
set $\mathfrak{P}_{C}^{<}$ are labeled as left-moving poles, and
those in set $\mathfrak{P}_{C}^{>}$ as right-moving poles. The number
of elements in a set $\mathfrak{S}$ is denoted by $|\mathfrak{S}|$.

On the other hand, symmetries can relate the Green's function at different positions in the BZ, leading to band degeneracy. Consequently, symmetries can impose constraints on the properties of the Green's function. In this section, we consider time-reversal symmetry $\mathcal{T}\equiv U_{T}\mathcal{K}$
($\mathcal{T}^{2}=-1$), where $\mathcal{K}$ denotes complex conjugate,
as a specific example to illustrate its influence on the properties of the Green's function.

The time-reversal symmetry dictates that the Green's function satisfies
the equation \citep{PhysRevResearch.5.033043}: 
\begin{equation}
U_{T}G\left(z,\mathbf{k}\right)^{t}U_{T}^{\dagger}=G\left(z,-\mathbf{k}\right),\label{eq:TR}
\end{equation}
where $U_{T}=-U_{T}^{t}$ represents the unitary part of the time-reversal
operator and $t$ denotes transposition. This symmetry denoted as
type C symmetry of Class 7 in Ref. \citep{zhou2019periodic}, which
guarantees that for each eigenstate of $G$, there is an associated
pair with the same complex eigenvalue, providing robust Kramers degeneracy
at time-reversal invariant momenta. Incorporating this symmetric Green's function, we propose the following proposition:

\textit{Proposition:} If the loop $C$ in the BZ is a time-reversal-invariant
closed line, e.g., $k_{j\neq x}=0$ and $k_{x}\in\left[-\pi,\pi\right)$
in the BZ, then $|\mathfrak{Z}_{C}^{<}|-|\mathfrak{Z}_{C}^{>}|-|\mathfrak{P}_{C}^{<}|+|\mathfrak{P}_{C}^{>}|=0$.
\begin{proof}
The antisymmetric unitary matrix of the antiunitary time reversal
symmetry operator implies that the dimension of the Green's function
is even. Additionally, for the type C symmetry of Class 7, the generalized
Kramers degeneracy ensures that all eigenvalues of the Green's function
$G\left(i0^{+},\mathbf{k}\right)$ come in pairs with opposite chiralities\citep{zhou2019periodic}.
Therefore, one has:

\begin{equation}
|\mathfrak{Z}_{C}^{<}|-|\mathfrak{Z}_{C}^{>}|-|\mathfrak{P}_{C}^{<}|+|\mathfrak{P}_{C}^{>}|=\ointop_{C}d\mathbf{k}\cdot\nabla_{\mathbf{k}}\ln\det G(i0^{+},\mathbf{k})=\ointop_{C}d\mathbf{k}\cdot\nabla_{\mathbf{k}}\sum_{j=1}^{j=Dim}\ln\lambda_{j}(i0^{+},\mathbf{k})=0,\label{eq:lndetG}
\end{equation}
where $Dim$ and $\lambda_{j}\left(i0^{+},\mathbf{k}\right)$ are
the dimension and eigenvalues of the Green's function $G\left(i0^{+},\mathbf{k}\right)$,
respectively. 
\end{proof}
It's worth noting that in the above proof, we don't assume the system
to be noninteracting. Therefore, the restriction on the number of
zeros and poles of the Green's function holds true for interacting
many-body Fermionic systems with time-reversal symmetry as well.

\section{Recursive methods for the surface Green's
function}
\label{app:F}

In this section, we provide a brief overview of the recursive Green's
function used to obtain the surface Green's function in the main text.
More details can be found in the reference (\citep{sancho1985highly}).

For a solid with a surface, it can be described by a semi-infinite
stack of layers with nearest-neighbor inter-layer interaction. Without
loss of generality, let's set the boundary at z=0, and the bulk extends
in the positive z direction. The lattice periodicity on the (x, y)-plane
is preserved. Thus, $\mathbf{k}=(k_{x},k_{y})$ is a good quantum
number, and each layer can be described by the Bloch wavefunctions
$\psi_{n}(\mathbf{k})$ with n denoting the layer indices. Taking
matrix elements of $(z-H)G=I$ between the Bloch wavefunctions, one
has the equations for each $\mathbf{k}$: 
\begin{equation}
\begin{cases}
(z-H_{00})G_{0,0} & =I+H_{01}G_{1,0}\\
(z-H_{00})G_{i,0} & =H_{01}^{\dagger}G_{i-1,0}+H_{01}G_{i+1,0};i\in\mathbf{Z}^{+},
\end{cases}\label{eq:gg1}
\end{equation}
where the matrices 
\begin{equation}
\begin{cases}
H_{nm} & =\braket{\psi_{n}|H|\psi_{m}}\\
G_{n,m} & =\braket{\psi_{n}|G|\psi_{m}},
\end{cases}
\end{equation}
and $I$ is the unit matrix. Since each layer is described by the
same Hamiltonian and the same coupling between nearest neighbor layers,
one has $H_{00}=H_{11}=\ldots$ and $H_{01}=H_{12}=\ldots$. From
Eq. (\ref{eq:gg1}), we get

\begin{equation}
\begin{cases}
(z-E_{s})G_{0,0} & =I+\alpha_{1}G_{2,0}\\
(z-E_{1})G_{2i,0} & =\beta_{1}G_{2\left(i-1\right),0}+\alpha_{1}G_{2\left(i+1\right),0}\\
(z-E_{1})G_{2i,2i} & =I+\beta_{1}G_{2\left(i-1\right),2i}+\alpha_{1}G_{2\left(i+1\right),2i},
\end{cases}\label{eq:gg3}
\end{equation}
where 
\begin{equation}
\begin{cases}
\alpha_{1} & =H_{01}(z-H_{00})^{-1}H_{01}\\
\beta_{1} & =H_{01}^{\dagger}(z-H_{00})^{-1}H_{01}^{\dagger}\\
E_{1} & =H_{00}+H_{01}(z-H_{00})^{-1}H_{01}^{\dagger}+H_{01}^{\dagger}(z-H_{00})^{-1}H_{01}\\
E_{s} & =H_{00}+H_{01}(z-H_{00})^{-1}H_{01}^{\dagger}.
\end{cases}\label{eq:gg4}
\end{equation}
Equations (\ref{eq:gg3}) define a chain that couples the Green's
function matrix elements with even indices only. Equations (\ref{eq:gg4})
define an effective Hamiltonian describing a chain with a lattice
constant twice the original lattice, and the effective intra-layer
and inter-layer hopping are renormalized. Starting from Eq. (\ref{eq:gg3})
and repeating the same procedure n times, we get 
\begin{equation}
G_{s}=G_{0,0}\approx(z-E_{s,n})^{-1}
\end{equation}
where $E_{s,n}$ is obtained by performing the following iteration
n times, ensuring that $\alpha_{n}$ and $\beta_{n}$ can be made
as small as desired, 
\begin{equation}
\begin{cases}
E_{s,j} & =E_{s,j-1}+\alpha_{j-1}(z-E_{j-1})^{-1}\beta_{j-1}\\
\alpha_{j} & =\alpha_{j-1}(z-E_{j-1})^{-1}\alpha_{j-1}\\
\beta_{j} & =\beta_{j-1}(z-E_{j-1})^{-1}\beta_{j-1}\\
E_{j} & =E_{j-1}+\alpha_{j-1}(z-E_{j-1})^{-1}\beta_{j-1}+\beta_{j-1}(z-E_{j-1})^{-1}\alpha_{j-1},
\end{cases}
\end{equation}
where $E_{0}=H_{00}$, $\alpha_{0}=H_{0,1}$ and $\beta_{0}=H_{0,1}^{+}$.
After n iterations, the surface layer is equivalent to the original
surface layer coupled to $2^{n}$ layers. 

\section{ Reconstruction of surface Green's function}
\label{app:G}

{\blue{In this section, we use a lattice model in the topological insulating phase as an example to illustrate the reconstruction of the surface Green's function $G_{s}(\omega+i 0^{+}, \mathbf{k})$ with components $G^{\alpha_i\alpha_j}_{s}(\omega+i 0^{+}, \mathbf{k})$, where $\alpha_i$ and $\alpha_j$ label the spin indices. We emphasize that although this method can, in principle, reconstruct the surface Green's function, it is challenging to implement with current techniques. For example, spin-resolved ARPES data \cite{neupane2016observation} provides diagonal components of the imaginary part of the surface Green's function; however, to our knowledge, the off-diagonal components have not yet been observed experimentally.

Firstly, we employ the recursive method to obtain the surface Green's function and extract its imaginary part, which constitutes the spectral function $A\left(\omega,\mathbf{k}\right)=\frac{-1}{\pi}Im G_{s}\left(\omega+i 0^{+},\mathbf{k}\right)$ that can be approximated by ARPES data.
Secondly, the Kramers-Kronig relations: 
$Re G_{s}\left(\omega+i 0^{+},\mathbf{k}\right)= P \int_{-\infty}^{\infty} d \Omega\frac{A\left(\Omega,\mathbf{k}\right)}{\omega-\Omega}$, are utilized to obtain the real part of the surface Green's function and where $P$ denotes the Cauchy principal value. Generally, a larger number of spectral function values across frequencies is required to accurately determine the real part of the surface Green's function, but in principle, this is feasible.
Finally, by combining the real part of the surface Green's function with the ARPES data, we reconstruct the complete surface Green's function $G_{s}^{\text{Rec}}\left(\omega+i 0^{+},\mathbf{k}\right) = Re G_{s}\left(\omega+i 0^{+},\mathbf{k}\right) -i \pi A\left(\omega,\mathbf{k}\right)$.

In Table I, we present the minimum value of the singular values of the reconstructed surface Green's function $G_{s}\left(i 0^{+},\mathbf{0}\right)$ for the TI, where the surface Dirac cone is gapped by a surface perturbation breaking time-reversal symmetry.
The integration in the Kramers-Kronig relations is performed from -3 to 3, which is larger than the bandwidth of the lattice model, and the integration step is $2\times10^{-7}$. The deviation of singular values of the reconstructed surface Green's function from the singular values of the actual surface Green's function indicates that a larger number of data points of the spectral function are needed to obtain a faithfully reconstructed surface Green's function.}}
\begin{table}[h]
    \centering
    \begin{tabular}{|c|c|c|c|c|}
        \hline
        $ \text{SVD}\left(G_{s}\left(i0^{+},\mathbf{0}\right)\right)$ & 5 & 5 &  $10^{-6}$&$10^{-6}$\\[6pt]
        \hline
        \rowcolor{white} 
        ${\text{SVD}\left(G_{s}^{\text{Rec}}\left(i0^{+},\mathbf{0}\right)\right)}$ &4.98 & 4.98&  0.018&0.018\\[6pt]
        \hline
    \end{tabular}
    \caption{Singular values of the \emph{surface} Green's function $G_{s}\left(i0^{+},\mathbf{0}\right)$ and the reconstruction of the \emph{surface} Green's function $G_{s}^{\text{Rec}}\left(i0^{+},\mathbf{0}\right)$ from the imaginary part of the $G_{s}\left(i0^{+},\mathbf{0}\right)$. }
    \label{tab:my_table}
\end{table}

\end{document}